\documentclass[aps,twocolumn,pre]{revtex4}
\usepackage{graphicx}
\usepackage{latexsym}
\usepackage{amsmath}
\usepackage{amsfonts}
\usepackage{amssymb}
\usepackage{color}

\begin{document}
\title{Electrophoretic Capture of a DNA Chain into a Nanopore}
\author{Payam Rowghanian}
\email{payam.rowghanian@physics.nyu.edu}
\author{Alexander Y. Grosberg}
\affiliation{Department of Physics and Center for Soft Matter
Research, New York University, 4 Washington Place, New York, NY
10003, USA}
\date{\today}

\begin{abstract}
Based on our formulation of the DNA electrophoresis near a pore [P. Rowghanian and A. Y. Grosberg, Phys. Rev. E 87, 042723 (2013)], we address the electrophoretic DNA capture into a nanopore as a steady-state process of particle absorption to a sink placed on top of an energy barrier. Reproducing the previously observed diffusion-limited and barrier-limited regimes as two different limits of the particle absorption process and matching the data, our model suggests a slower growth of the capture rate with the DNA length for very large DNA molecules than the previous model, motivating more experiments beyond the current range of electric field and DNA length. At moderately weak electric fields, our model predicts a different effect, stating that the DNA length dependence of the capture rate first disappears as the field is reduced and eventually reverses to a decreasing trend with $N$.
\end{abstract} 

\maketitle

\section{Introduction}

Driven electrophoretically into a nanopore drilled in a membrane, DNA molecules exhibit a complex behavior in translocation experiments \cite{Kasianowicz26111996}. Numerous works have focused on the threading stage of the translocation process \cite{branton2008potential}, and another handful have studied the capture of the DNA into the pore before it begins to pass through \cite{de1999passive,wanunu2009electrostatic,grosberg2010dna,muthukumar2010theory}. Experiments \cite{wanunu2009electrostatic} have observed an initial significant increase followed by saturation of the capture rate with DNA length. The capture rate monotonously increases with the applied voltage, but the character of this increase is different for weak and strong applied voltages.

An explanation of the observed features was suggested in Refs.  \cite{wanunu2009electrostatic,grosberg2010dna} with the participation of one of us (A. Y. G.). Although
successful in several respects, it involved at one point a tentative \textit{ad hoc} assumption of nearly complete suppression of electro-osmotic flow through the DNA coil when it came to the proximity of the membrane. Following up on this point, we have improved this consideration in our accompanying work \cite{PayamShuraElectro} and shown that while considerable, the suppression of electro-osmosis by the membrane is far from complete. The main purpose of the present work is now to examine the implications of that finding for the interpretation of the capture experiments \cite{wanunu2009electrostatic}. We show that the correctly calculated suppression of electro-osmosis is sufficient to explain the data and that it also provides a detailed account of possible regimes and their crossover. In addition, our theory yields a prediction that upon lowering the applied voltage, the DNA length dependence of the capture rate first disappears and eventually reverses so that the capture becomes a decreasing function of the DNA length.

After a short description of the translocation experiment, in the first part of this article, we present a reminder of our accompanying work regarding the DNA electrophoresis for a coil placed near a pore and of previous works regarding the capture theory \cite{wanunu2009electrostatic,grosberg2010dna}. We then revisit the capture problem by viewing the distance of the DNA end from the pore entrance as a single relevant reaction coordinate. Although the applicability of this reaction coordinate is obvious while the coil is far from the pore and can be treated as a point-like object, we argue that it remains marginally applicable up until the DNA end touches the pore. We then compute the free energy landscape which is determined by the entropy of a DNA coil near a membrane as well as its electrophoretic pull near and far from a pore. The two different regimes of the capture process introduced in Refs. \cite{wanunu2009electrostatic,grosberg2010dna} naturally appear as two limits of this consideration.

\section{Background}

\subsection{Experimental Setup}

\begin{figure}
 \includegraphics[width=0.95\linewidth]{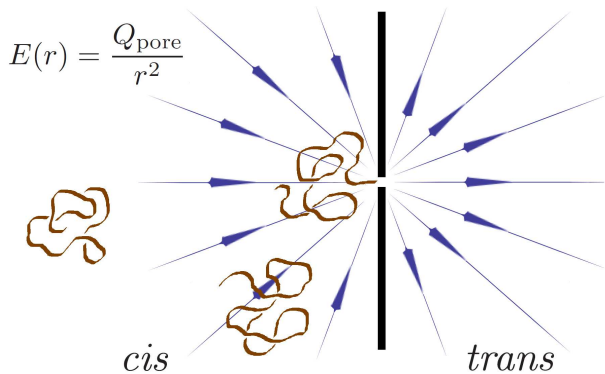}
 \caption{Schematic representation of DNA translocation through a nanopore. A dilute electrolyte solution of DNA molecules is subject to an external voltage, which produces an electric field $E(r)$ inside the electrolyte and electrophoretically drives the DNA molecules towards and through the pore. The electric field is like one produced by a point charge $Q_\mathrm{pore}$ which is determined by the pore dimensions and the applied voltage, is almost uniform along the DNA molecules far from the pore, and is highly non-uniform for the molecules close to or captured in the pore. \label{setup}}
\end{figure}

The setup of a translocation experiment is schematically shown in Figure \ref{setup}. An electrolyte dilute solution of DNA molecules is separated by a dielectric membrane into a \emph{cis} and \emph{trans} part. Upon the application of an external voltage to the apparatus, the charged DNA molecules are drawn towards a very nanometer-sized pore drilled in the membrane, and pass through the pore from the \emph{cis} to the \emph{trans} side of the membrane. The electric field which drives this process is not subject to Debye screening, as it is maintained by an electric current driven by the voltage through the conducting medium and the pore and in this sense, it is the result of a steady but not equilibrium process. 

As shown in ref. \cite{wanunu2009electrostatic}, far enough from the apparatus walls, the electric field has a spherical geometry and is like that of a point charge, with a potential $V(r) =  Q_{\mathrm{pore}}/r$ as a function of the distance $r$ from the pore, valid at $r>a$, where $a$ is the pore width. The effective ``pore charge'' is $Q_{\mathrm{pore}} = \Delta V a^2/(8b)$, where for simplicity, the total applied voltage $\Delta V$ sits in the place of the voltage drop across the pore which depends linearly on $\Delta V$. The non-uniform field $E(r)=Q_{\mathrm{pore}}/r^2$ created by the applied voltage facilitates the DNA capture into the pore by electrophoretically attracting the DNA towards the pore; this effect, which we have studied in detail in our accompanying work \cite{PayamShuraElectro} is summarized below. 

\subsection{DNA electrophoresis near a nanopore}

Biasing the ion concentrations, the contour of a DNA coil is effectively surrounded by a thin layer of positively charged liquid, which opposes the DNA motion upon the application of an electric field $E$. This reduces the effective electrophoretic pull, which is often characterized using the force necessary to stall the DNA against the electric field
\begin{equation}
F_{\mathrm{st}}=Q_{\mathrm{eff}} E.\label{stall force}
\end{equation}
The stall force $F_{\mathrm{st}}$ depends on $E$ through an effective charge $Q_{\mathrm{eff}}=\mu_E/\mu_F$, where $\mu_F$ is the standard mechanical mobility and $\mu_E\sim \lambda \ln{\left(1+ {r_D}/{d}\right)}/\eta$ is the well-known length independent electrophoretic mobility \cite{cleland1991electrophoretic,muthukumar1996theory,PhysRevLett.76.3858}, in which, $d$ and $\lambda$ are the DNA width and charge per unit chain length, $\eta$ is the solvent viscosity and $r_D$ is the electric screening radius, understood often in the context of Debye theory. Using the linear relation between the electro-osmotic flow and the electric field, and the same $1/r^2$ decay of the electric and velocity field with the distance $r$ from the pore, we have obtained in \cite{PayamShuraElectro} the stall force per DNA segment for a coil captured from one end into the pore to be
\begin{equation}
 f_{st}(r) = q_{\mathrm{eff}}(r) E(r) \sim \mu_E \eta\ell \left(\frac{r}{\ell}\right)^\frac{\nu-1}{\nu} E(r),\label{stall force per segment}
\end{equation}
where $\ell$ is the Kuhn segment length and the effective charge per segment is $q_{\mathrm{eff}}(r)=\mu_E/\mu_F(r)$, with $\mu_F(r)$ formally viewed as the mechanical mobility of a single segment at $r$, inversely proportional to the friction coefficient per segment $\xi(r)\sim \eta \ell (r/\ell)^{(\nu-1)/\nu}$ in a blob of size $r$. The stall force in Eq (\ref{stall force per segment}) characterizes the viscous suppression (by electro-osmotic flow) of the bare electric pull; its increase at small $r$ indicates the natural weakening of the counterflow in the vicinity of the membrane. Equation (\ref{stall force per segment}) is valid for voltages which are not strong enough to significantly deform the DNA; this corresponds to $u\sim 1$ or smaller, with the dimensionless parameter $u$ defined as
\begin{equation}
u=\frac{ \eta\ \mu_E Q_{\mathrm{pore}}}{T}, \label{parameter u}
\end{equation}
representing the strength of the electric field relative to thermal energy $T$. This turns out to be the conditions under which experiments \cite{wanunu2009electrostatic} have been performed. For $u\gg 1$, the DNA is highly compressed and concentration blobs form (see our accompanying work \cite{PayamShuraElectro} for a detailed discussion). 

\subsection{Summary of the previous DNA capture theory}

Here we briefly summarize the theoretical ideas of the works \cite{wanunu2009electrostatic,grosberg2010dna} that we use here to highlight the places where improvements are necessary. 

\emph{Non-interacting DNA molecules:} For a sufficiently dilute solution of DNA molecules, each molecule is captured into the pore independently of the others and therefore, the capture rate is liner in the bulk DNA concentration $c_0$ and is equal to $R_c c_0$. It is this quantity $R_c$ that is usually measured in experiments, which for brevity is called the capture rate. We will consider the same quantity in our analysis below.

\emph{Quasiequilibrium DNA energy landscape:} The way the DNA capture is facilitated by the electric field can be formulated in terms of its energy landscape in the potential field $V(r)$ introduced above. Assuming that the DNA does not contribute significantly to the electric current which maintains the electric field $E(r)$, its presence does not affect the potential $V(r)$. Under this assumption, the DNA can be considered as moving in a potential field $-W(r)$, where 
\begin{equation}
 W(r) = \int_{\infty}^{r} F_{\mathrm{st}}(r') \mathrm{d}r'  \simeq \eta\ \mu_E Q_{\mathrm{pore}}\frac{R}{r} \label{w formula}
\end{equation} 
is the work of the stall force on the DNA coil as it is delivered quasistatically from infinity to the distance $r$ from the pore and can be written as $W(r) = Q_{\mathrm{eff}} V(r)$. Equation (\ref{w formula}) is only valid for $r \gg R$ with $R$ being the coil size, where field variations across the DNA coil are small and thus the coil can be considered as a point-like particle.

\emph{Diffusion versus barrier limited regimes:} Capture of a DNA coil into a pore occurs through an interplay between the DNA diffusion far from the pore and the electrophoretic DNA drift close to the pore, and takes place in two different regimes. In the ``barrier-limited'' regime, the DNA molecules which arrive at the pore face an entropic barrier, which takes many attempts to overcome and is the rate limiting factor in the capture rate. In the ``diffusion-limited'' regime, the entropic barrier is mostly flattened by the electrophoretic pull and the limiting factor is the rate at which DNA molecules arrive diffusively to a capture radius $r^*$ from the pore.

\emph{Diffusion-limited regime:} For sufficiently large applied voltages and DNA lengths, the rate limiting process is the diffusive arrival of the DNA coils to a capture radius $r^*$ introduced below, in which case, the process can be described by the classical Smoluchowski theory for diffusive particle absorption \cite{smoluchowski1917versuch} with a rate
\begin{equation}
R_c^{\mathrm{diff}} = 2 \pi \mathcal{D} r^{\ast},\label{diffusion limited original} 
\end{equation}
where $R_c^{\mathrm{diff}}$ is the diffusion-limited rate, $\mathcal{D}$ is the coil diffusion coefficient, and the coefficient $2\pi$ instead of the familiar $4\pi$ appears because the DNA is captured in a half space only. The capture radius $r^{\ast}$, the distance at which DNA free diffusion at large distances crosses over to the electrophoretic drift down the potential $V(r)$, can be found from the condition $W(r^{\ast}) = T$ (using the units in which the Boltzmann constant equals $1$ and $T$ is the thermal energy) to be
\begin{equation}
r^*\sim \frac{Q_{\mathrm{pore}}\mu_E}{\mathcal{D}}, \label{capture radius}
\end{equation}
where $\mu_E$ is the DNA length independent electrophoretic mobility. This yields a DNA length independent diffusion-limited capture rate $R_c^{\mathrm{diff}}$ for large DNA length $N$. Clearly, this result, based on considering the DNA coil as a point-like particle, is only valid as long as $r^{\ast} \gg R$, where $R$ is the coil size. For the parameters of the experiment \cite{wanunu2009electrostatic}, it was estimated \cite{grosberg2010dna} that the ratio $r^{\ast}/R \approx 2$, which means the theory is at the border of applicability. The ratio $r^{\ast}/R$ is the same as the dimensionless electric field parameter $u$ introduced in the next section (Eq (\ref{parameter u})).

\emph{Barrier-limited regime:} In the opposite limit, the rate limiting process is overcoming the pore entrance barrier. In this regime, the DNA coil arrives many times at the pore entrance before it finally succeeds in placing its end into the pore and beginning to thread. Therefore, there is almost an equilibrium between the DNA at the pore and in the bulk. For the DNA to introduce its end into the pore, the coil has to overcome a free-energy barrier and, therefore, the capture process in this regime can be viewed as overcoming a barrier, whose rate, using Kramers theory \cite{kramers1940brownian,hanggi1990reaction}, is
\begin{equation}
 R_c^{\mathrm{bar}}=\omega\exp\left(-\frac{F_b}{T}\right),\label{Kramers formula}
\end{equation}
where $\omega$ is the attempt rate (with units $\mathrm{nM}^{-1} \mathrm{s}^{-1}$) and $F_b$ is the barrier height. The barrier height decreases with the electric field. This decrease is of a subtle nature, as when the DNA is at the pore, it is subject to a very non-uniform electric field and the electro-osmotic flow is affected by the proximity of the membrane. To describe this, a tentative \textit{ad hoc} assumption was made in \cite{wanunu2009electrostatic,grosberg2010dna} that $Q_{\mathrm{eff}}$ in this situation was close to the bare DNA charge and thus linear in the DNA length $N$. Although this allowed to explain the increase of the capture rate with $N$ for moderate values of $N$, the arbitrary character of this assumption called for improvement. This in fact is the main purpose of an accompanying work \cite{PayamShuraElectro}, where we show that the assumption $Q_{\mathrm{eff}} \sim N$ while sufficient to account for the existing data, is far too bold and results in an overestimation of the DNA energy at the pore.

Other aspects of the capture theory, namely, the size-dependent conformational and orientational entropic cost of bringing the DNA end into the pore, and a clear characterization of the crossover between the barrier-limited and diffusion-limited regimes, were also not considered in the previous works; they will be considered in the present work.

\section{DNA Capture into the Pore\label{capture section}}

\subsection{Smoluchowski particle absorption process}

Taking the same approach as in \cite{grosberg2010dna}, we model the problem as a steady state particle absorption process, in which, DNA molecules are provided far from the pore with a constant flux $-J=c_0 R_c$, with $c_0$ the DNA concentration far from the pore and $R_c$ the capture rate, and are absorbed as they get captured into the pore, through which, the same flux $-J$ of molecules passes. Any successful capture event involves the arrival of a DNA molecule to a position of low enough free energy with respect to its starting points far from the pore, which makes that event practically an irreversible event and the pore a sink for the DNA molecules. The rate at which the DNA molecules are captured (into the sink, or the pore), as pointed out by \citeauthor{smoluchowski1917versuch} \cite{smoluchowski1917versuch}, is characterized by the concentration profile of those molecules; far from the pore, where the DNA can be considered as a point-like particle, the DNA concentration $c(r)$ satisfies the Smoluchowski equation \cite{smoluchowski1917versuch}
\begin{equation}
c_0 R_c= -J=\mathcal{D}r^2\left(\nabla c(r)+\frac{c(r)}{T}\nabla F(r)\right),\label{total current}
\end{equation}
where $F(r)$ is the free energy landscape. The applicability of Eq (\ref{total current}) is obvious far from the pore, where the coil can be viewed as a point-like object and $r\gg R$ can be the distance of any part of the coil from the pore entrance. When the coil approaches the pore, we must specify what variable to use as a reaction coordinate to account for the dynamics of the system.  We found the following very simple idea useful. Let us define $r$ as the distance of the DNA end from the pore entrance; we ignore for the time being the fact that there are two ends. This choice of coordinate is a reasonable and simple single-variable measure of how close to the state of being captured from one end the coil is; since the relaxation time of the end as it diffuses a distance comparable to the coil size within the coil remains of the same order as the relaxation time of the coil, $r$ is marginally applicable as a single reaction coordinate which approximately describes how the different states between a coil merely sitting at the pore and captured from one end are explored by the coil. For the same reason and as another manifestation of this marginal applicability, we assume that the effective diffusion constant remains unchanged as the system evolves along $r$.  This assumption is certainly not exact, but remains at the margin of applicability as the end approaches the pore, and so must yield the right scaling results.

The steady state absorption problem formulated using the Smoluchowski Equation (\ref{total current}) must be  equipped with an absorbing boundary condition $c(r_b)=0$, imposed at the barrier peak $r_b \sim \ell$ which corresponds to the DNA end touching the pore. Imposing this condition, we find 
\begin{equation}
c(r)=c_0 \frac{R_c}{\mathcal{D}}\exp{\left(-\frac{F(r)}{T}\right)}\int_{r_b}^{r}\exp{\left(\frac{F(r')}{T}\right)} \frac{ \mathrm{d}r'}{{r'}^2},
\end{equation}
which can be used to obtain the rate $R_c$ by letting the total number of particles in the system of volume $V$ be equal to $c_0 V$ and thus
\begin{equation}
\frac{V\mathcal{D}}{R_c}=\int_{r_b}^{\infty} \exp{\left(-\frac{F(r)}{T}\right)}\left[\int_{r_b}^{r}\exp{\left(\frac{F(r')}{T}\right)} \frac{ \mathrm{d}r'}{{r'}^2}\right]\mathrm{d}r.\label{total particles}
\end{equation}

\subsection{Derivation of the free energy landscape}

\begin{figure}
 \includegraphics[width=0.99\linewidth]{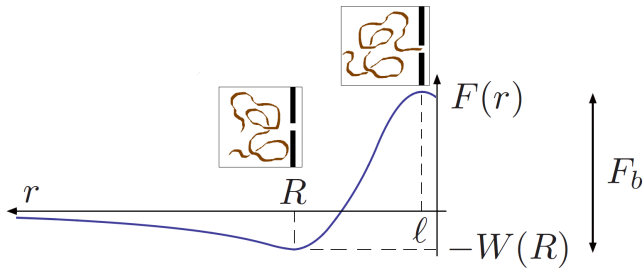}
 \caption{DNA free energy landscape $F(r)$ in the barrier-limited regime as a function of the reaction coordinate $r$, set to be the position of the DNA end with respect to the pore. For $r>R$, where $R$ is the DNA coil size, the free energy is solely determined by the coil's electrophoretic attraction (Eq (\ref{w formula})) and decreases to $-W(R)$ as the coil is brought to a distance $R$ from the pore. For $r<R$, and as one DNA end moves from within the coil towards the pore, the conformational and orientational entropy of the coil also contribute to the free energy; for the barrier-limited regime, this contribution results in an overall increase $F_b$ in the free energy as the DNA end is brought from a distance $\sim R$ to $\sim \ell$, which is the height of the free energy barrier. The free energy begins to fall again for $r\lesssim \ell$, as the strong electric field inside the pore performs work on the DNA segments which pass through the pore. \label{free energy}}
\end{figure}

In this subsection, we calculate the free energy landscape $F(r)$ of the DNA as a function of the reaction coordinate $r$, which consists of an electrophoretic part and an entropic cost imposed on a DNA with one end held in the pore. Here is an overall glimpse of $F(r)$, as also sketched in Fig. \ref{free energy}. Starting from zero far from the pore, $F(r)=-W(r)$ [Eq. (\ref{w formula})] decreases smoothly for $r>R$ and reaches a local minimum $-W(R)$ as it arrives at the pore. From this minimum, one DNA end can be brought to the pore. Such a motion would involve an energetic gain $W_p(r)$ and an entropy cost $S_p(r)$ and therefore, the free energy would be $F(r)=-W(R)-T S_p(r)-W_p(r)$ for $r<R$. As we will discuss in the end in more details, $F$ reaches its maximum at $r_b\sim \ell$, which is the top of the free energy barrier and corresponds to one DNA end held near the pore.

The next few paragraphs are devoted to the derivation of the free energy near the pore, i.e. for the reaction coordinate $r<R$. Let us first calculate the entropy $S_p$ of introducing one DNA end into the pore. $S_p=S_p^{\mathrm{con}}+S_p^{\mathrm{or}}$ contains a conformational term $S_p^{\mathrm{con}}$ and a term $S_p^{\mathrm{or}}$ related to the orientational freedom of the end. The conformational term appears when the DNA end is brought to a distance $\sim \ell$ from the pore; during this step, the conformational freedom of the coil is reduced to that of a flexible polymer grafted from one end to a solid membrane. The resulting entropic cost is $S_p^{\mathrm{con}}(\ell)\sim (g/\nu)\ln{(\ell/R)}$, where the constant $g=\gamma-\gamma_s$ is related to the entropic exponents appearing while counting the number of polymer conformations far from ($\gamma$) and near ($\gamma_s$) a surface \cite{des1990polymers}. More generally, bringing the DNA to a distance $r\ll R$ from the pore results in $S_p^{\mathrm{con}}(r)\sim (g/\nu)\ln{(r/R)}$; this relation could be understood by rescaling the segment size from $\ell$ to $r$.

The coil end must be oriented in such a way that it can enter the pore; this orientation begins to occur when the first DNA segment (counted from its end) is at a distance $\ell$ from the pore and costs an amount $S_p^{\mathrm{or}}$. Let us remember that the Kuhn length $\ell$ for dsDNA is much larger than the pore dimensions and therefore, entering the pore results in a rotational restriction of the captured end, reducing the total $4\pi$ solid angle available to a free end to an amount $\sim (a/b)^2$, where $a$ and $b$ are pore dimensions. This results in an entropy cost $S_p^{\mathrm{or}}\sim \ln{(4\pi b^2/a^2)}$. The change of the reaction coordinate upon the entrance of the coil end into the pore is very small and on the order of the microscopic length scale $\ell$ of the coil. 

We now obtain the energetic part of the free energy $W_p(r)$ by first calculating the work of the stall force as one DNA end is brought from within the coil to a distance $\sim \ell$ from the pore. For this to happen, the coil is pulled in such a way that any segment indexed $g$ with respect to the captured end is brought to a distance $r\sim \ell g^\nu$ from the pore, required for a polymer grafted from one end to a surface \cite{vanderzande1998lattice,grimmett2001probability}. During this motion, on average, $\left(r/\ell\right)^{(1-\nu)/(\nu)} ({\mathrm{d}r }/{\ell})$ segments are brought from a distance $\sim R$ to $r$ from the pore and placed in a shell of thickness $\mathrm{d}r$. This motion occurs while liquid flows through and around the coil and therefore, the work of stall force performed on a single segment brought from $R$ to $r$ is
\begin{equation}
 w_{\mathrm{seg}}(r)\sim\int_{R}^r f_{\mathrm{st}}(r') \mathrm{d}r' \sim \eta\ \mu_E Q_{\mathrm{pore}} \left(\frac{\ell}{r}\right)^{\frac{1}{\nu}}
\end{equation}
where $f_{\mathrm{st}}(r')$ is found by substituting $E(r')=Q_{\mathrm{pore}}/r'^2$ in Eq (\ref{stall force per segment}). Summing over all the segments, we obtain 
\begin{equation}
 W_p(\ell)\sim\int_\ell^{R} w_{\mathrm{seg}}(r) \left(\frac{r}{\ell}\right)^\frac{1-\nu}{\nu} \frac{\mathrm{d}r }{\ell} \sim \eta\ \mu_E Q_{\mathrm{pore}} \ln{\frac{R}{\ell}}.\label{wb formula}
\end{equation}
A scaling argument similar to the one used above for obtaining $S_p^{\mathrm{con}}(r)$ will suggest here that $W_p(r) \sim \eta\ \mu_E Q_{\mathrm{pore}} \ln{(R/r)}$ for $r\ll R$. Let us note that the electrophoretic mobility is independent of segment size; therefore, since the integration in Eq. (\ref{wb formula}) is dominated by the upper bound and thus the energetic gain is determined just like the entropic loss by the largest scale of the coil, rescaling the segment size from $\ell$ to $r$ only affects the argument inside the logarithm.

Putting now the entropic cost $T S_p$ and $W_p$ together, the free energy landscape as a function of the reaction coordinate far from and near the pore is
\begin{subequations}\begin{align}
 \frac{F(r<R)}{T}=&\frac{g}{\nu}\ln{\frac{R}{r}}+\Theta_\ell(r)\ln{\frac{4\pi b^2 }{a^2}}-u \left(\frac{w_p}{\nu}\ln{\frac{R}{r}}+w\right), \label{capture free energy}\\
 \frac{F(r>R)}{T}=&-wu\frac{R}{r}, 
 \end{align}\end{subequations}
where $\Theta_\ell(r)$ is $1$ for $r\lesssim \ell$ and $0$ for $r\gtrsim \ell$, the constant $g=(\gamma-\gamma_s)>0$ corresponds to the conformational entropy cost, and $w$ and $w_p$ are positive numerical factors appearing in $W(R)=wuT$ and $W_p(\ell)=(w_p/\nu) u T\ln(R/\ell)$ with the dimensionless effective electric field $u$ defined in Eq (\ref{parameter u}); we have included $\nu$ explicitly in $W_p(\ell)$ as a convention for brevity of the expressions below. The barrier height $F_b=F(\ell)-F(R)$ can then be found as
\begin{equation}
 \frac{F_b}{T}=-\epsilon \ln{\frac{R}{\ell}}+\ln{\frac{4\pi b^2}{a^2}}, \label{free energy barrier}
\end{equation}
where $\epsilon=(w_p u-g)/\nu$ represents the electrophoretic $W_p$ and conformational $T S_p^{\mathrm{con}}$ parts combined; $\epsilon=0$ corresponds to the case in which the conformational entropy cost of the end capture is fully compensated for by the electrophoretic pull. As we will see later when we compare our results with experimental data, the conformational and electrophoretic parts make a small contribution to the barrier and the main contribution comes from the orientational term [the last term in Eq. (\ref{free energy barrier})].

We end this part with two comments about the barrier top which we have proposed to be at $r_b\sim \ell$. First, at this level of approximation where the coil dynamics is described by a single reaction coordinate, it cannot be determined with certainty at what reaction coordinate the first segment of the coil begins to lose its orientational entropy. However, since a capture attempt in which the first segment comes very close (such that $r\ll \ell$) to the pore sideways is very likely to be rejected, it is reasonable to propose that the dominant paths in the capture process are the ones in which the first segment is fairly aligned with the pore axis when it is not closer than a distance $\sim \ell$ from the pore. Also, further beyond this point and when the tip of the first segment enters the pore, as we show in our accompanying work \cite{PayamShuraElectro}, an extra amount of energy $\sim \eta\mu_E Q_{\mathrm{pore}} \ln{(\ell/a)}$ is gained by the coil; this is the work of the stall force [Eq. (\ref{stall force per segment})] as the tip of one segment is brought inside the pore, and it determines how much the free energy drops as the reaction coordinate $r$ decreases from $\sim \ell$ to $\sim a$. Both of these observations hint at the plausibility of the proposition that the free energy is peaked around a point $r_b\sim \ell$.

Second, after the DNA end is captured and $n$ segments have passed through the pore, the coil continues to lose more conformational entropy, known to be $\gamma_s T[\ln{(N-n)}+\ln{n}]$ \cite{sung1996polymer}. This entropic cost, however, is overcome by the energetic gain of threading which scales linearly with $n$ and is $\sim n\ \eta \mu_E Q_{\mathrm{pore}}$ (assuming the field range used in experiments and neglecting the extra complication regarding electrophoresis in a pore as well as the length independent electrostatic cost of holding a DNA segment inside the pore \cite{zhang2007effective}). As a result, the free energy beyond the barrier peak $r_b$ decreases and thus, the absorbing boundary condition at the barrier peak is correctly imposed.

\subsection{Diffusion-limited and barrier-limited regimes}

\begin{figure}
 \includegraphics[width=0.95\linewidth]{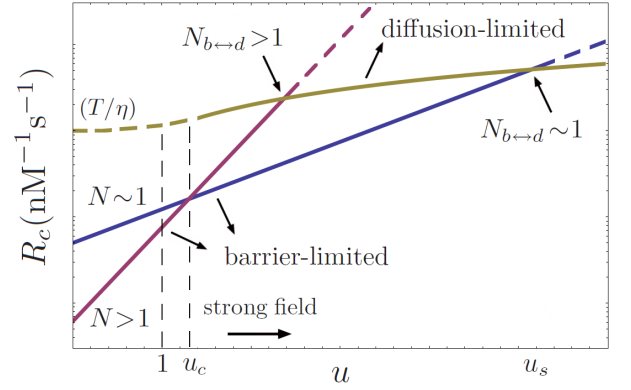}
 \caption{A sketch of the capture rate in logarithmic scale as a function of the dimensionless field strength $u=\eta\ \mu_E Q_{\mathrm{pore}}/T$. The curved line on the top corresponds to the diffusion-limited rate $R_c^{\mathrm{diff}}$, which is independent of the DNA length $N$. Its value is $\sim (T/\eta)$ [Eq. (\ref{weak field diffusion-limited rate})] for weak fields $u\lesssim 1$ and increases linearly as $(T/\eta)u$ [Eq. (\ref{diffusion-limited rate})] for strong fields $u\gtrsim 1$. The two straight lines correspond to the barrier-limited rate $R_c^{\mathrm{bar}}$ for $N=1$ and $N>1$. The barrier-limited rate grows exponentially with $u$.  It decreases with $N$ for weak fields [Eq. (\ref{weak field barrier-limited rate})], but at $u=u_c\sim 1$, this trend reverses and for strong fields, longer DNA molecules have a higher barrier-limited rate [Eq (\ref{barrier-limited rate})]. The barrier-limited curve crosses over to the diffusion-limited curve as the field is further increased. The field strength at which this crossover from barrier-limited to diffusion-limited regimes takes place is larger for smaller DNA molecules. At a large enough field $u=u_s$, the crossover happens at $N_{b\leftrightarrow d}\sim 1$, which means that above $u_s$, diffusion is the rate limiting factor for all DNA lengths. The experimentally observed capture rate at any given $u$ is determined by the lowest of the barrier- and diffusion-limited ones; some parts of each curve which will never be observable in experiments for any $N$ have been drawn as dashed lines to emphasize this point. \label{Rc schematic}}
\end{figure}

We now derive $R_c$ by substituting the free-energy landscape into Eq. (\ref{total particles}). This consideration is aimed at finding $R_c$ in different regimes, as summarized by Eqs. (\ref{weak field barrier-limited rate}), (\ref{weak field diffusion-limited rate}), (\ref{barrier-limited rate}) and (\ref{diffusion-limited rate}) and sketched schematically in Fig. \ref{Rc schematic}. To calculate the integrals in Eq (\ref{total particles}), we break the one over $r'$ into two parts. The first part $\mathcal{I}_1$ runs from $\ell$ to $R$, and the second one $\mathcal{I}_2(r)$ runs from $R$ to $r$ (since the integral over $r$ scales with the experimental apparatus volume $V$ and is thus dominated by $r\to \infty$, we will not calculate the inner integral for $r<R$). We obtain the second integral to be
\begin{equation}
\mathcal{I}_2(r)=\frac{1}{wuR}\left(e^{-wu\frac{R}{r}}-e^{-wu}\right). \label{second inner integral}
\end{equation}
The first one depends on $\epsilon=(w_p u-g)/\nu$ and is 
\begin{subequations}\begin{align}
\mathcal{I}_1\approx& \frac{1}{\ell} e^{-wu+\phi_b}+\frac{1}{R}e^{-wu}&, \ \ \ \epsilon>1\label{inner integral 1}\\
\mathcal{I}_1\approx& \frac{1}{\ell} e^{-wu+\phi_b}&, \ \ \ \epsilon<1\label{inner integral 2}
\end{align}\end{subequations}
where we have dropped a small term of order $a^2/(4\pi b^2)$ in both Eqs (\ref{inner integral 1}) and (\ref{inner integral 2}) and kept the dominant terms (the last term in Eq (\ref{inner integral 1}) only begins to become important when $\epsilon>1$). Substituting now Eqs (\ref{second inner integral}), (\ref{inner integral 1}) and (\ref{inner integral 2}) into Eq (\ref{total particles}) we obtain
\begin{subequations}\begin{align}
\frac{\mathcal{D}}{R_c}\approx&\ \frac{e^{-wu}}{R}\left(\frac{R}{\ell}e^{\phi_b} +1\right)+\frac{1}{wuR}\left(1-e^{-wu}\right)\ , \ \epsilon>1, \label{inverse rate epsilon>1}\\
\frac{\mathcal{D}}{R_c}\approx&\ \frac{1}{\ell}e^{-wu+\phi_b}+\frac{1}{wuR}\left(1-e^{-wu}\right)\ , \ \epsilon<1. \label{inverse rate epsilon<1}
\end{align}\end{subequations}

Equations (\ref{inverse rate epsilon>1}) and (\ref{inverse rate epsilon<1}) are relations for inverse current or overall resistance, in which, the overall resistance $\mathcal{D}/R_c=\Omega_b+\Omega_d$ is found as the sum of a barrier resistor and a diffusion resistor connected in series. $\Omega_b$ corresponds to the resistance molecules face when they attempt to overcome the barrier and $\Omega_d$ corresponds to the bulk resistance molecules face as they diffusively arrive at the pore. For weak fields $u\ll 1$, the barrier resistance $\Omega_b=\exp{(-wu+\phi_b)}/\ell$ in Eq (\ref{inverse rate epsilon<1}) dominates and results in the barrier-limited rate
\begin{equation}
 R_{\mathrm{weak}}^{\mathrm{bar}}\sim \frac{T}{\eta} \frac{a^2}{4\pi b^2} N^{-(g+\nu)}\ \ , \ \ u\ll 1,\label{weak field barrier-limited rate}
\end{equation}
which decays with $N$ due to both slower diffusion of larger molecules towards the pore ($N^{-\nu}$) and the growth of the entropic barrier with the DNA size ($N^{-g}$). Although overshadowed by the barrier term, we could also formally pick only the diffusion resistance $\Omega_d=\left(1-e^{-wu}\right)/(wuR)$ to find
\begin{equation}
 R_{\mathrm{weak}}^{\mathrm{diff}}\sim \frac{T}{\eta} \ \ , \ \ u\ll 1,\label{weak field diffusion-limited rate}
\end{equation}
which as expected, is much larger than $R_{\mathrm{weak}}^{\mathrm{bar}}$ and thus not observed in experiments as the rate is limited by the barrier term. $R_{\mathrm{weak}}^{\mathrm{diff}}$ does not depend on the electric field at all and smoothly crosses over at $u\sim 1$ to the field-dependent diffusion-limited rate for strong fields [Eq. (\ref{diffusion-limited rate}) below]. From the point of view of the approach used in \cite{wanunu2009electrostatic}, this can be interpreted as the capture radius $r^*$ [Eq. (\ref{capture radius})] smoothly crossing over to $R$ for weak fields $u< 1$. 

The behavior of the capture rate for moderately strong fields [Eq. (\ref{inverse rate epsilon>1})] is slightly more complex. On the one hand, the two parts of the barrier resistance $\Omega_b={e^{-wu}}\left(Re^{\phi_b}/\ell +1\right)/R$ compete with each other and on the other hand, they both compete with the diffusion resistance $\Omega_d=\left(1-e^{-wu}\right)/(wuR)$. As a result of this, we can identify two crossover DNA lengths, namely, $N_{b\leftrightarrow d}$, where the diffusion term dominates, and $N^*$, where the two terms of the barrier term become comparable. We find 
\begin{equation}
 N^* \sim \left(\frac{4\pi b^2}{a^2}\right)^\frac{1}{\nu(\epsilon-1)}, \label{crossover N*}
\end{equation}
above which, $\Omega_b\approx e^{-wu}/R\ll \Omega_d$. Below $N^*$, the barrier resistance is $\Omega_b\approx \exp{(-wu+\phi_b)}/\ell$, which becomes comparable to the diffusion resistance at
\begin{equation}
 N_{b\leftrightarrow d}\sim \left(\frac{4\pi b^2}{a^2}\frac{u}{e^{wu} \left(1-e^{-wu}\right)}\right)^\frac{1}{\nu(\epsilon-1)}. \label{barrier diffusion crossover}
\end{equation}
Using Eq (\ref{barrier diffusion crossover}), we can now find the barrier-limited and diffusion-limited capture rates for a moderately strong field $u\gtrsim 1$ to be 
\begin{align}
 R_{\mathrm{str}}^{\mathrm{bar}}\sim&\ \frac{T}{\eta}\ e^{wu} \frac{a^2}{4\pi b^2} N^{(-\nu+w_p u-g)}&,\ \ N<N_{b\leftrightarrow d},\label{barrier-limited rate}\\
 R_{\mathrm{str}}^{\mathrm{diff}}\sim&\ \frac{T}{\eta}u &,\ \ N>N_{b\leftrightarrow d}.\label{diffusion-limited rate}
\end{align}
The experimentally observed capture rate corresponds to Eqs (\ref{barrier-limited rate}) and (\ref{diffusion-limited rate}) for $N<N_{b\leftrightarrow d}$ and $N>N_{b\leftrightarrow d}$ respectively. The diffusion-limited rate, in agreement with the earlier theory and experiment \cite{wanunu2009electrostatic}, is independent of the DNA size. The barrier-limited rate increases with a power $\alpha= -\nu+w_p u-g$ of $N$ for strong enough fields or for $u>(\nu+g)/w_p\sim 1$, which matches the experimental data \cite{wanunu2009electrostatic} while being functionally different from the stretched exponential form obtained previously \cite{wanunu2009electrostatic}. Rewriting Eq. (\ref{barrier-limited rate}) in the form of Eq (\ref{Kramers formula}) will result in an attempt rate $\omega\sim (T/\eta) N^{-\nu}$, which has an extra $N^{-\nu}$ term in comparison to the value $\omega\sim (T/\eta)$ in \cite{wanunu2009electrostatic}. 

In addition to the increasing capture rate with $N$ for moderately strong fields, our solution predicts a different trend, stating that as the field is decreased to pass through the value $u_c= (\nu+g)/w_p$, the $N$-dependence of the barrier-limited rate first disappears and then reverses so that the capture rate begins to decay with $N$. For weak fields, the rate eventually crosses over to the purely diffusive rate [Eq. (\ref{weak field barrier-limited rate})], which may result in too rare capture events to be experimentally measurable; however, the crossover region $u\sim u_c$ and moderately weak fields may be feasible to observe experimentally. 

The diffusion-limited to barrier-limited crossover length $N_{b\leftrightarrow d}$ decreases with $u$ for moderately strong fields. Eventually, at a strong enough field $u_s$, we obtain $N_{b\leftrightarrow d}\sim 1$ and therefore, the capture process becomes diffusion-limited for all DNA lengths. The capture rate as a function of $u$ in different regimes and for two different values of $N$ is schematically plotted in Fig. \ref{Rc schematic}.

\section{Comparison with Experiments}

Below we compare our results with experiments \cite{wanunu2009electrostatic} and obtain the quantities of interest up to numerical factors of order unity; we find the dimensionless field strength to be $u\gtrsim 1$, the screening radius to be ${r_D}\sim{d}\sim 1 \mathrm{nm}$, and the crossover DNA length to be $N_{b\leftrightarrow d}\approx 20$. Note that all the quantities obtained here are valid only as order of magnitude estimations to demonstrate the consistency of our model with experimental data.

\begin{figure}
 \includegraphics[width=0.96\linewidth]{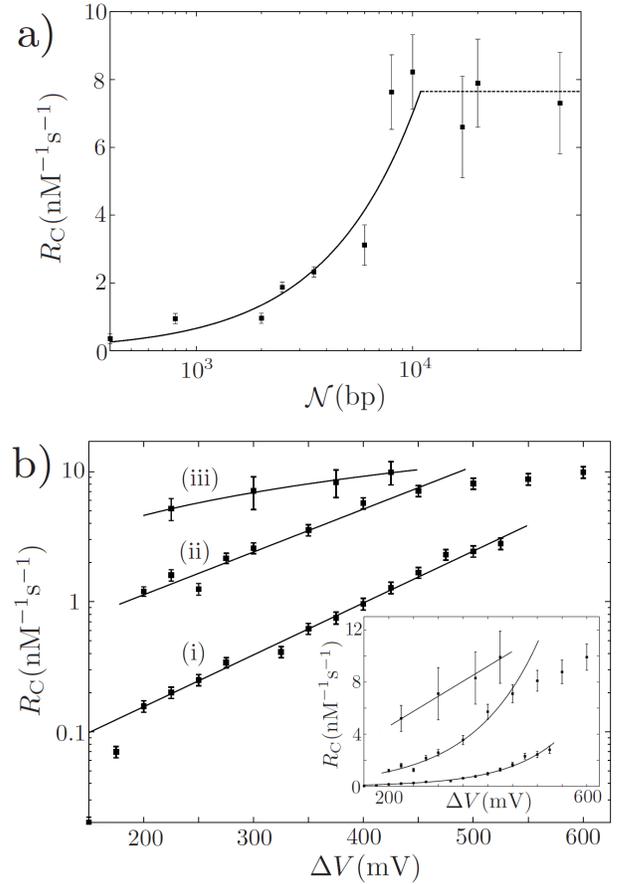}
 \caption{a) Capture rate as a function of DNA length $\mathcal{N}$ for $\Delta V=300\mathrm{mV}$. The solid line corresponds to the barrier-limited capture rate $R_{\mathrm{str}}^{\mathrm{bar}}= A \mathcal{N}^\alpha$, with $\alpha=-\nu+w_p u-g$ and $A=\left(n_{\mathrm{Kuhn}}\right)^{\nu+g-w_p u} (T/\eta) e^{wu}\left[a^2/(4\pi b^2)\right]$ (Eq (\ref{barrier-limited rate})), where $u=\eta\ \mu_E Q_{\mathrm{pore}}/T$, and $g=\gamma-\gamma_s$ is related to the surface exponents of the coil far from ($\gamma$) and near ($\gamma_s$) a membrane, with the numerical value $g= 0.5$ for an ideal chain, which is the configuration of the polymers for the lengths used in these experiments. From the fit we get $\alpha=1.03\pm 0.16$ and $A\approx 6\times 10^{-4} \mathrm{nM}^{-1} \mathrm{s}^{-1}$; using the value of $T/\eta = 2.4 \mathrm{nM}^{-1} \mathrm{s}^{-1}$ and $A$, we find that the crossover to the diffusion-limited regime (dashed line) occurs at $\mathcal{N}_{b\leftrightarrow d}\approx 6000\mathrm{bp}$, consistent with the observed value $\mathcal{N}_{b\leftrightarrow d} \approx 9000\mathrm{bp}$. The dashed line corresponds to the diffusion-limited rate $R_c^{\mathrm{diff}}= \mu_E Q_{\mathrm{pore}}\approx  8 \mathrm{nM}^{-1} \mathrm{s}^{-1}$, which using the value of $\lambda Q_{\mathrm{pore}}/\eta\approx 6 \mathrm{nM}^{-1} \mathrm{s}^{-1}$ gives $\ln{\left(1+{r_D}/{d}\right)}\approx 1.3$. b) Capture rate as a function of applied voltage $\Delta V$ for $\mathcal{N}_{(i)}=400 \mathrm{bp}$, $\mathcal{N}_{(ii)}=3500 \mathrm{bp}$, and $\mathcal{N}_{(iii)}=48000 \mathrm{bp}$. Starting from the top, the data set (iii) corresponds to the diffusion-limited regime, in which $R_{\mathrm{str}}^{\mathrm{diff}}=D \Delta V$ (Eq (\ref{diffusion-limited voltage dependence})), with $D=\mu_E a^2/(8b)$. From the fit we obtain $D\approx 23 \mathrm{nM}^{-1} \mathrm{s}^{-1} \mathrm{V}^{-1}$, which using ${\lambda a^2}/(8b \eta)\approx 18 \mathrm{nM}^{-1} \mathrm{s}^{-1} \mathrm{V}^{-1}$ reproduces $\ln{\left(1+{r_D}/{d}\right)}\approx 1.3$. The data sets (i) and (ii) were assumed to correspond to the barrier-limited regime. Although the data seems to fit to $R_{\mathrm{str}}^{\mathrm{bar}}= B \exp(C \Delta V)$ (Eq (\ref{barrier-limited rate})), the fitted values of $B$ for (i) and (ii) show an increase rather than decrease with $N$. We interpret this as a sign that these data belong in fact to the crossover rather than the barrier-limited region, which is indeed the trend seen in the last three data points of (ii). \label{Rc graph}}
\end{figure}

Figure \ref{Rc graph}(a) shows our results fitted to the experimental data \cite{wanunu2009electrostatic} for the capture rate as a function of the DNA size $\mathcal{N}=n_{\mathrm{Kuhn}}N$ in basepairs, with $n_{\mathrm{Kuhn}}=300 \mathrm{bp}$ being the length of a Kuhn segment. The dashed line corresponds to the $N$ independent diffusion-limited regime (Eq (\ref{diffusion-limited rate})) and indicates that $R_c^{\mathrm{diff}}\sim \mu_E Q_{\mathrm{pore}}\approx 8  \mathrm{nM}^{-1} \mathrm{s}^{-1}$. Using the experimental values $\Delta V=300\mathrm{mV}$, pore dimensions $a=5\mathrm{nm}$ and $b=25\mathrm{nm}$, the Manning threshold $\lambda\approx 1.5\ \mathrm{e}/\mathrm{nm}$, and $\eta=10^{-3}\mathrm{Pa}. \mathrm{s}$ for water, we find ${\lambda Q_{\mathrm{pore}}}/{\eta}\approx 6   \mathrm{nM}^{-1} \mathrm{s}^{-1}$. This then results in $\ln{\left(1+{r_D}/{d}\right)}\approx 1.3$. The solid line in Figure \ref{Rc graph}a corresponds to the barrier-limited rate (Eq (\ref{barrier-limited rate})). The fit indicates that the capture rate increases with a power $\alpha=-\nu+w_p u-g=1.03\pm 0.16$ of $N$ and also that
\begin{equation}
  \frac{T}{\eta} \frac{a^2}{4\pi b^2} e^{wu}\approx {0.2}\ \mathrm{nM}^{-1} \mathrm{s}^{-1},
\end{equation}
 which using ${T}/{\eta}\approx 2.4   \mathrm{nM}^{-1} \mathrm{s}^{-1}$ yields
\begin{equation}
 N_{b\leftrightarrow d} \approx 20, 
\end{equation}
in good agreement with the experiments, where crossover appears to occur around $\mathcal{N}_{b\leftrightarrow d}\approx 9000$. 

Taking $\Delta V$ as the independent variable, we can rewrite the capture rate (Eqs (\ref{barrier-limited rate}) and (\ref{diffusion-limited rate})) as
\begin{equation}
 R_{\mathrm{diff}}\sim \frac{\lambda a^2}{8b\eta}\ln{\left(1+\frac{r_D}{d}\right)}  \Delta V,  \label{diffusion-limited voltage dependence}
\end{equation}
\begin{equation}
\begin{split}
 R_c^{\mathrm{bar}}\sim &\ \frac{T}{\eta} \left(\frac{a^2}{4\pi b^2}\right) N^{-(g+\nu)}\times \\ &\exp\left[\ln{\left(1+\frac{r_D}{d}\right)} \frac{\lambda a^2}{8b T} (w+w_p\ln N) \Delta V\right]. \label{barrier-limited voltage dependence}
\end{split}
\end{equation}
Figure \ref{Rc graph}(b) shows the experimental data \cite{wanunu2009electrostatic} for the capture rate as a function of the applied voltage. The data set with $N_{\mathrm{(iii)}}\approx 160$ falls in the diffusion-limited regime [Eq. (\ref{diffusion-limited voltage dependence})]. The linear fit gives $\ln{\left(1+{r_D}/{d}\right)}\sim 1.3$, where we have used the value of $\lambda a^2/(8b T)\approx 6 \mathrm{V}^{-1}$. The two data sets with $N_{\mathrm{(i)}}\approx 1$ and $N_{\mathrm{(ii)}}\approx 12$ seem to correspond to the barrier-limited regime [Eq. (\ref{barrier-limited voltage dependence})] with $R_c^{\mathrm{bar}}= B\exp(C \Delta V)$. The constants $B$ and $C$ obtained from the fit seem to exhibit the opposite of the expected trends, i.e. $B$ seems to increase rather than decrease with $N$, and $C$ does not quite show an increase with $\ln{N}$ and slightly decreases with $N$. While the latter might be associated with the fact that the logarithmic growth really corresponds to very large values of $N$ (much larger than $N_{\mathrm{(ii)}}\approx 12$), the reverse trend of $B$ cannot be ignored. We explain this ostensible discrepancy by stating that we believe the data indeed are collected in a region where a crossover from the barrier-limited to the diffusion-limited regime is taking place; in fact, the last few points in the data set (ii) show a clear deviation from the exponential trend to the linear trend, characteristic of the diffusion-limited regime. Therefore, the data are significantly bent and the values obtained from the fits must not be taken seriously. Using Eq (\ref{barrier diffusion crossover}), indeed one can show that the voltage at which $N_{b\leftrightarrow d}\sim 1$ corresponds to $u\approx 6$, which is about $900 \mathrm{mV}$, not very far from the range at which data are taken. 

\section{Conclusion}

Using the formulation of the electrophoresis of a DNA coil at the pore, we develop a concise model of the capture process in a translocation experiment, which produces the previously observed diffusion-limited and barrier-limited regimes as two different limits separated by a crossover naturally emerging from the solution. While matching the data, our model suggests a much slower growth of the barrier-limited capture rate with the DNA length than the previous model \cite{wanunu2009electrostatic}, which will be observable for very large DNA molecules, beyond the range at which the barrier-limited regime has been experimentally observable so far. The experimental challenge for testing this is that for very large $N$, the system crosses over to the diffusion-limited regime, in which, the capture rate is independent of the DNA length. One way to overcome this difficulty would be to increase the crossover size $N_{b\leftrightarrow d}$ by lowering the electric field (while keeping the exponent $\alpha$ just above zero) or by using a deeper pore with a larger ratio $b/a$; the effectiveness of both methods, however, is limited. As an alternative, our model may be tested experimentally by exploring moderately weak electric fields, at which, the DNA length dependence of the barrier-limited capture rate is expected to first disappear and eventually reverse to a decreasing trend with $N$. 

\begin{acknowledgments}
This research was supported in part by the National Science Foundation under Grant No. NSF PHY11-25915. We would like to thank the Kavli Institute for Theoretical Physics in Santa Barbara where part of this work was done. The work of A. Y. G. was supported in part by a grant from the U.S.-Israel Binational Science Foundation, and P.R. was supported by the National Science Foundation under Grant No. NSF PHY-0424082. We would like to also acknowledge constructive discussions with Y. Rabin, B. Shklovskii, and L. Lizana.
\end{acknowledgments}


\begin{thebibliography}{18}
\expandafter\ifx\csname natexlab\endcsname\relax\def\natexlab#1{#1}\fi
\expandafter\ifx\csname bibnamefont\endcsname\relax
  \def\bibnamefont#1{#1}\fi
\expandafter\ifx\csname bibfnamefont\endcsname\relax
  \def\bibfnamefont#1{#1}\fi
\expandafter\ifx\csname citenamefont\endcsname\relax
  \def\citenamefont#1{#1}\fi
\expandafter\ifx\csname url\endcsname\relax
  \def\url#1{\texttt{#1}}\fi
\expandafter\ifx\csname urlprefix\endcsname\relax\def\urlprefix{URL }\fi
\providecommand{\bibinfo}[2]{#2}
\providecommand{\eprint}[2][]{\url{#2}}

\bibitem[{\citenamefont{Kasianowicz et~al.}(1996)\citenamefont{Kasianowicz,
  Brandin, Branton, and Deamer}}]{Kasianowicz26111996}
\bibinfo{author}{\bibfnamefont{J.~J.} \bibnamefont{Kasianowicz}},
  \bibinfo{author}{\bibfnamefont{E.}~\bibnamefont{Brandin}},
  \bibinfo{author}{\bibfnamefont{D.}~\bibnamefont{Branton}}, \bibnamefont{and}
  \bibinfo{author}{\bibfnamefont{D.~W.} \bibnamefont{Deamer}},
  \bibinfo{journal}{Proc. Natl. Acad. Sci. (U.S.A.)}
  \textbf{\bibinfo{volume}{93}}, \bibinfo{pages}{13770} (\bibinfo{year}{1996}).

\bibitem[{\citenamefont{Branton et~al.}(2008)\citenamefont{Branton, Deamer,
  Marziali, Bayley, Benner, Butler, Di~Ventra, Garaj, Hibbs, Huang
  et~al.}}]{branton2008potential}
\bibinfo{author}{\bibfnamefont{D.}~\bibnamefont{Branton}},
  \bibinfo{author}{\bibfnamefont{D.~W.} \bibnamefont{Deamer}},
  \bibinfo{author}{\bibfnamefont{A.}~\bibnamefont{Marziali}},
  \bibinfo{author}{\bibfnamefont{H.}~\bibnamefont{Bayley}},
  \bibinfo{author}{\bibfnamefont{S.~A.} \bibnamefont{Benner}},
  \bibinfo{author}{\bibfnamefont{T.}~\bibnamefont{Butler}},
  \bibinfo{author}{\bibfnamefont{M.}~\bibnamefont{Di~Ventra}},
  \bibinfo{author}{\bibfnamefont{S.}~\bibnamefont{Garaj}},
  \bibinfo{author}{\bibfnamefont{A.}~\bibnamefont{Hibbs}},
  \bibinfo{author}{\bibfnamefont{X.}~\bibnamefont{Huang}},
  \bibnamefont{et~al.}, \bibinfo{journal}{Nature Biotechnology}
  \textbf{\bibinfo{volume}{26}}, \bibinfo{pages}{1146} (\bibinfo{year}{2008}).

\bibitem[{\citenamefont{de~Gennes}(1999)}]{de1999passive}
\bibinfo{author}{\bibfnamefont{P.-G.} \bibnamefont{de~Gennes}},
  \bibinfo{journal}{Proc. Natl. Acad. Sci.} \textbf{\bibinfo{volume}{96}},
  \bibinfo{pages}{7262} (\bibinfo{year}{1999}).

\bibitem[{\citenamefont{Wanunu et~al.}(2009)\citenamefont{Wanunu, Morrison,
  Rabin, Grosberg, and Meller}}]{wanunu2009electrostatic}
\bibinfo{author}{\bibfnamefont{M.}~\bibnamefont{Wanunu}},
  \bibinfo{author}{\bibfnamefont{W.}~\bibnamefont{Morrison}},
  \bibinfo{author}{\bibfnamefont{Y.}~\bibnamefont{Rabin}},
  \bibinfo{author}{\bibfnamefont{A.}~\bibnamefont{Grosberg}}, \bibnamefont{and}
  \bibinfo{author}{\bibfnamefont{A.}~\bibnamefont{Meller}},
  \bibinfo{journal}{Nature Nanotech.} \textbf{\bibinfo{volume}{5}},
  \bibinfo{pages}{160} (\bibinfo{year}{2009}).

\bibitem[{\citenamefont{Grosberg and Rabin}(2010)}]{grosberg2010dna}
\bibinfo{author}{\bibfnamefont{A.}~\bibnamefont{Grosberg}} \bibnamefont{and}
  \bibinfo{author}{\bibfnamefont{Y.}~\bibnamefont{Rabin}}, \bibinfo{journal}{J.
  Chem. Phys.} \textbf{\bibinfo{volume}{133}}, \bibinfo{pages}{165102}
  (\bibinfo{year}{2010}).

\bibitem[{\citenamefont{Muthukumar}(2010)}]{muthukumar2010theory}
\bibinfo{author}{\bibfnamefont{M.}~\bibnamefont{Muthukumar}},
  \bibinfo{journal}{J. Chem. Phys.} \textbf{\bibinfo{volume}{132}},
  \bibinfo{pages}{195101} (\bibinfo{year}{2010}).

\bibitem[{\citenamefont{Rowghanian and Grosberg}(2013)}]{PayamShuraElectro}
\bibinfo{author}{\bibfnamefont{P.}~\bibnamefont{Rowghanian}} \bibnamefont{and}
  \bibinfo{author}{\bibfnamefont{A.~Y.} \bibnamefont{Grosberg}},
  \bibinfo{journal}{Phys. Rev. E} \textbf{\bibinfo{volume}{87}},
  \bibinfo{pages}{042723} (\bibinfo{year}{2013}).

\bibitem[{\citenamefont{Cleland}(1991)}]{cleland1991electrophoretic}
\bibinfo{author}{\bibfnamefont{R.}~\bibnamefont{Cleland}},
  \bibinfo{journal}{Macromolecules} \textbf{\bibinfo{volume}{24}},
  \bibinfo{pages}{4391} (\bibinfo{year}{1991}).

\bibitem[{\citenamefont{Muthukumar}(1996)}]{muthukumar1996theory}
\bibinfo{author}{\bibfnamefont{M.}~\bibnamefont{Muthukumar}},
  \bibinfo{journal}{Electrophoresis} \textbf{\bibinfo{volume}{17}},
  \bibinfo{pages}{1167} (\bibinfo{year}{1996}).

\bibitem[{\citenamefont{Long et~al.}(1996)\citenamefont{Long, Viovy, and
  Ajdari}}]{PhysRevLett.76.3858}
\bibinfo{author}{\bibfnamefont{D.}~\bibnamefont{Long}},
  \bibinfo{author}{\bibfnamefont{J.-L.} \bibnamefont{Viovy}}, \bibnamefont{and}
  \bibinfo{author}{\bibfnamefont{A.}~\bibnamefont{Ajdari}},
  \bibinfo{journal}{Phys. Rev. Lett.} \textbf{\bibinfo{volume}{76}},
  \bibinfo{pages}{3858} (\bibinfo{year}{1996}).

\bibitem[{\citenamefont{von Smoluchowski}(1917)}]{smoluchowski1917versuch}
\bibinfo{author}{\bibfnamefont{M.}~\bibnamefont{von Smoluchowski}},
  \bibinfo{journal}{Z. Phys. Chem} \textbf{\bibinfo{volume}{92}},
  \bibinfo{pages}{129} (\bibinfo{year}{1917}).

\bibitem[{\citenamefont{Kramers}(1940)}]{kramers1940brownian}
\bibinfo{author}{\bibfnamefont{H.}~\bibnamefont{Kramers}},
  \bibinfo{journal}{Physica} \textbf{\bibinfo{volume}{7}}, \bibinfo{pages}{284}
  (\bibinfo{year}{1940}).

\bibitem[{\citenamefont{H{\"a}nggi et~al.}(1990)\citenamefont{H{\"a}nggi,
  Talkner, and Borkovec}}]{hanggi1990reaction}
\bibinfo{author}{\bibfnamefont{P.}~\bibnamefont{H{\"a}nggi}},
  \bibinfo{author}{\bibfnamefont{P.}~\bibnamefont{Talkner}}, \bibnamefont{and}
  \bibinfo{author}{\bibfnamefont{M.}~\bibnamefont{Borkovec}},
  \bibinfo{journal}{Rev. Mod. Phys.} \textbf{\bibinfo{volume}{62}},
  \bibinfo{pages}{251} (\bibinfo{year}{1990}).

\bibitem[{\citenamefont{Des~Cloizeaux and Jannink}(1990)}]{des1990polymers}
\bibinfo{author}{\bibfnamefont{J.}~\bibnamefont{Des~Cloizeaux}}
  \bibnamefont{and} \bibinfo{author}{\bibfnamefont{G.}~\bibnamefont{Jannink}},
  \emph{\bibinfo{title}{Polymers in Solution: Their Modelling and Structure}}
  (\bibinfo{publisher}{Clarendon Press Oxford}, \bibinfo{year}{1990}).

\bibitem[{\citenamefont{Vanderzande}(1998)}]{vanderzande1998lattice}
\bibinfo{author}{\bibfnamefont{C.}~\bibnamefont{Vanderzande}},
  \emph{\bibinfo{title}{{Lattice Models of Polymers}}}
  (\bibinfo{publisher}{Cambridge University Press, Cambridge},
  \bibinfo{year}{1998}).

\bibitem[{\citenamefont{Grimmett and
  Stirzaker}(2001)}]{grimmett2001probability}
\bibinfo{author}{\bibfnamefont{G.}~\bibnamefont{Grimmett}} \bibnamefont{and}
  \bibinfo{author}{\bibfnamefont{D.}~\bibnamefont{Stirzaker}},
  \emph{\bibinfo{title}{{Probability and Random Processes}}}
  (\bibinfo{publisher}{Oxford University Press, New York},
  \bibinfo{year}{2001}).

\bibitem[{\citenamefont{Sung and Park}(1996)}]{sung1996polymer}
\bibinfo{author}{\bibfnamefont{W.}~\bibnamefont{Sung}} \bibnamefont{and}
  \bibinfo{author}{\bibfnamefont{P.~J.} \bibnamefont{Park}},
  \bibinfo{journal}{Phys. Rev. Lett.} \textbf{\bibinfo{volume}{77}},
  \bibinfo{pages}{783} (\bibinfo{year}{1996}).

\bibitem[{\citenamefont{Zhang and Shklovskii}(2007)}]{zhang2007effective}
\bibinfo{author}{\bibfnamefont{J.}~\bibnamefont{Zhang}} \bibnamefont{and}
  \bibinfo{author}{\bibfnamefont{B.~I.} \bibnamefont{Shklovskii}},
  \bibinfo{journal}{Phys. Rev. E} \textbf{\bibinfo{volume}{75}},
  \bibinfo{pages}{021906} (\bibinfo{year}{2007}).

\end{thebibliography}

\end{document}